\documentclass[12pt]{article}
\usepackage{hyperref}
\usepackage{graphicx}
\usepackage{rotating}

\usepackage{amssymb}

 \def\a{\alpha}

\def\g{\gamma}

\def\d{\delta}
\def\e{\varepsilon}

\def\beq{\begin{equation}}
\def\eeq{\end{equation}}
\def\beqn{\begin{eqnarray}}
\def\eeqn{\end{eqnarray}}
\def\ba{\begin{eqnarray}}
\def\ea{\end{eqnarray}}

\def\m{{\tt -}}

\def\l{\langle}

\def\xprim2bar{\overline{x}^{\prime\prime}}

\def\beq{\begin{equation}}
\def\eeq{\end{equation}}

\newcommand{\beqa}{\begin{eqnarray}}
\newcommand{\eeqa}{\end{eqnarray}}

\let\a=\alpha      \let\g=\gamma   \let\d=\delta
\let\e=\epsilon         
      \let\l=\lambda  \let\m=\mu

\let\a=\alpha      \let\g=\gamma   \let\d=\delta
\let\e=\epsilon         
      \let\l=\lambda  \let\m=\mu

\newcommand{\be}{\begin{equation}}
\newcommand{\ee}{\end{equation}}
\newcommand{\bea}{\begin{eqnarray}}
\newcommand{\eea}{\end{eqnarray}}

\usepackage{graphics}
\begin{document}
\setcounter{page}{1}
\vspace{1.0cm}
\begin{center}{\large \bf Folding Froggatt-Nielsen into the St\"{u}ckelberg-Higgs mechanism in
anomalous $U(1)$ models}
\vspace{.17in}

\vspace{2cm}

{\bf\large Claudio Corian\`{o}$^{1}$ and $\;$  Nikos Irges$^{2}$}

\vspace{.12in}
\vspace{1cm}

{\it  $^1$Dipartimento di Fisica, Universit\`{a} di Lecce, and \\
INFN Sezione di Lecce,  Via Arnesano 73100 Lecce, Italy}\\
~\\
{\it
$^2$Department of Physics and Institute of Plasma Physics, \\
University of Crete, GR-710 03 Heraklion, Greece\\}
\end{center}
\vspace{.5cm}

\begin{abstract}
 
We describe a simple connection between the Froggatt-Nielsen, St\"{u}ckelberg and Higgs mechanisms, 
all three of them combined in a consistent way. This is illustrated in the context of 
a class of generalizations of the Standard Model with a gauge structure extended by a certain number of 
anomalous $U(1)$ factors. These are built in the effective action in a way that 
gauge invariance and unitarity are preserved. 
 Among other features, a physical axion with 
properties different from those of a Peccei-Quinn axion emerges.

\end{abstract}
\smallskip

\bigskip

{\bf Introduction.}
Understanding better mass hierarchy generating mechanisms is always 
a step forward in understanding nature. The one mechanism that we
think we understand best in this respect is the Higgs mechanism which generates masses
for the fermions and the gauge bosons of the Standard Model (SM).  
The hierarchy it creates in the gauge boson sector is not large but still very important
(it is one of the best measured physical properties in the SM) however since the Higgs couples in a 
universal way to the fermion families, it gives the same mass to all fermions.
In order to give the needed intricate flavor structure to the SM, the Froggatt-Nielsen
mechanism \cite{FNiel} can be employed, a version of which suggests the introduction of an 
extra abelian gauged $U(1)_X$ flavor symmetry along with a complex scalar singlet $\Theta$
that takes a vacum expectation value $<\Theta>\; =V$ and with the fermion mass hierarchy
being generated by higher dimensional operators in the $V$-vacum, suppressed by the
appropriate power of some high scale $M$.
The mechanism requires that the SM fermions are charged under $X$ and 
phenomenological viability of the mass matrices have the consequence that the extra $U(1)$
is "anomalous", i.e. summing over the fermions ($f$) gives non-vanishing $tr\{q^{(f)}_Xq^{(f)}_Xq^{(f)}_X\}$.
In an older analysis \cite{IL}, it was argued that a particularly fruitful scenario 
in models with multiple anomalous $U(1)$'s is when
the number of singlets is equal to the number of extra $U(1)$'s. 

In models where the $U(1)$ symmetries originate from a
string construction, the anomalies are naturally cancelled by the 
presence of appropriate axion couplings \cite{AK,CIK,ABDK,CIM1}, the four dimensional
version of the Green-Schwarz mechanism. Consistency of the theory 
requires also terms put in the effective Lagrangean 
in the original way St\"{u}ckelberg has suggested \cite{Stucky} with the axions playing the
role of the St\"{u}ckelberg auxiliary field. The effective Lagrangean
containing these terms is valid below $V$. 
At even lower energies, spontaneous symmetry breaking of the electroweak (EW) symmetry
occurs by the Higgs taking a vev $v$ and the Lagrangean must be rewritten in the new vacum,
the $v$-vacum. The hierarchy of scales in these models is $v<<V<M$.
Since many extensions of the SM involve more than one Higgs doublets,
including the MSSM and most string and D-brane inspired or derived models of this type \cite{IRU,AKT,IMR,AKRT,ADKS},
we will allow for such a possibility. In this case, in the scalar sector, an interesting interplay
between the St\"{u}ckelberg and Higgs mechanisms takes place \cite{CIK} where while in the
CP-even subsector things proceed pretty much as in the MSSM, in the CP-odd
subsector the axions mix with the corresponding Higgs components
\cite{Rub}, producing in the $v$-vacum
a new physical state, the axi-Higgs, denoted as $\chi$.  
In the vector sector, the combined St\"{u}ckelberg-Higgs mechanism generates a 
strong hierarchy of masses between the $Z'$-gauge boson and the EW gauge bosons and it
shifts by a small amount the (already small) hierarchy between the $Z$ and the $W$'s 
generated by the Higgs mechanism \cite{AK,GIIQ,G1}. 
The phenomenology of this sector has been looked at also in \cite{KN,FKN} whereas
generation of fermion mass hierarchies in this context has been attempted recently in \cite{CIM,CKL,GLR}.

Motivated by these considerations,
we propose an economical scheme where these three mechanisms are connected.
In fact, we will describe a mechanism where each extra $U(1)$ is accompanied a complex scalar field,
whose vev plays the role of a Froggatt-Nielsen vev, its phase plays the role of the 
St\"{u}ckelberg axion required for gauge invariance and unitarity at the quantum level
and upon EW symmetry breaking it (the phase) mixes with the CP-odd Higgs components
to form a number of NG-bosons but also the physical CP-odd state $\chi$ mentioned above.
 
{\bf The supersymmetric vacum}. We assume that just below $M$ the theory is supersymmetric.
The gauge group is $SU(3)\times SU(2)\times U(1)_Y\times U(1)_X$, 
$Y$ is hypercharge, $X$ is an extra $U(1)$, the
spectrum is that of the MSSM (plus right handed neutrinos perhaps) plus an extra singlet $\Theta$ and 
the fermion spectrum is anomalous with respect to $X$.
The charge of a field $\Phi$ under $U(1)_X$ is denoted as $q_X^{(\Phi)}$ and the corresponding
gauge boson as $A^X_\m$. 
The various couplings $g^{\Phi_1\Phi_2\Phi_3}$ with
$\Phi_{1,2,3}$ any fields have specific but rather complicated explicit expressions that can be found
in \cite{CIK,CIM1}. In the following we will be denoting a superfield and
its scalar component by the same letter;
the proper interpretation should be clear from the text.
With no loss of generality we assign the following $U(1)$ charges to the scalar sector:
%
\hskip 2cm
\begin{center}
\begin{tabular}{|c|c|c|}
\hline
$  $ & $Y$ &  $ X $  \\
\hline \hline
$\Theta$  &  $0$  & $q_X^{(\Theta)}$ \\ \hline
$H_u$    &  $1/2$  & $q_X^{(H_u)}$ \\ \hline
$H_d$    &  $1/2$  & $q_X^{(H_d)}$ \\ \hline
\end{tabular}
\end{center}
%
Supersymmetry dictates that in the $V$-vacum \cite{DSW}
(we neglect $F$-terms for the moment)
\bea
D_X &\sim& q_X^{(\Theta)} |\Theta|^2 + q_X^{(H_u)} |H_u|^2 + q_X^{(H_d)} |H_d|^2 - V^2 =0\nonumber\\
D_Y &\sim& |H_u|^2 + |H_d|^2 = 0\nonumber\\
D_{SU(2)} &\sim& |H_u^\dagger \tau^a H_u + H_d^\dagger \tau^a H_d|^2 = 0
\eea
where we have assumed that only $\Theta$ and $H_{u,d}$ can take a vev.  
Normalizing $q_X^{(\Theta)} = 1$ we obtain that in the $V$-vacum
\be
<|\Theta |> = V
\ee
and at this stage, even though $U(1)_X$ is broken to its global subgroup,
neither supersymmetry nor EW symmetry are broken.

{\bf St\"{u}ckelberg and anomalies}. Parametrizing as $\Theta = \rho\; e^{i \theta/M}$ and evaluating the 
$\Theta$ kinetic term in the $V$-vacum \cite{KNSusy} we obtain
\be
|D_\m \Theta|^2 = |\partial_\m \Theta + i A^X_\m \Theta|^2 = 
\frac{V^2}{M^2}(\partial_\m \theta + M A^X_\m)^2 + \cdots
\ee
which is just the familiar St\"{u}ckelberg coupling
that together with the dimension five term appearing in the expansion of the 
non-renormalizable coupling
\be
g^{\Theta F{\tilde F}} \frac{\Theta}{M} F\wedge F \longrightarrow {\rm total\; div.} +g^{\theta F{\tilde F}}\frac{1}{M}
 \;\theta F \wedge F + {\rm interactions}
\ee
and the anomaly, constitute the Green-Schwarz sector of the model. 
Notice that the scales $V$ and $M$ are not really independent, they are tied 
through the anomaly cancellation mechanism in a specific way.
The coupling $g^{\Theta F{\tilde F}}$ is constrained by gauge invariance \cite {CIK,ABDK} and unitarity \cite{CIM1}.
In the absence of a Higgs mechanism, the phase $\theta$ is the NG-boson associated
with the breaking of $X$ and there is a gauge where it is the longitudonal component of 
the massive $A^X_\m$. In the presence of the Higgs, one linear combination of 
$\theta$ and the CP-odd Higgs components becomes physical.
 
{\bf The Froggatt-Nielsen mechanism}. Let us now denote a SM invariant by $\cal S$. An example are the Yukawa couplings
${\bf Q}_i {\overline u}_j H_u$, etc. Clearly, if $\cal S$ is an invariant, $\frac{\Theta ^p}{M^p}\, {\cal S}$
is automatically $Y$-invariant but not necessarily $X$-invariant. The condition for 
the latter is $q_X^{(\cal S)} + p =0$.
In the $V$-vacum the invariant will appear as (putting explicit flavor indices $i,j=1,2,3$)
\be
\left(\frac{V}{M}\right)^{p_{ij}} {\cal S}_{ij}  + {\rm interactions},
\ee
i.e. it contains the usual Froggatt-Nielsen type of couplings that generate fermion mass
hierarchies when $X$ is flavor non-universal. Therefore from
now on we denote by ${V}/{M} = \l_c$,
the usual Froggatt-Nielsen expansion parameter which is chosen to be close in value to the Cabbibo angle
(in certain string constructions this is automatic).
What we have achieved is that by adding one additional parameter per extra $U(1)$ to a
St\"{u}ckelberg extension of the SM with anomalous $U(1)$'s, a mechanism that
generates fermion mass hierarchies is automatically present once a specific $X$-charge assignement is chosen
\cite{AF}.
 
{\bf Supersymmetry breaking}. Before we move to lower energies and to EW symmetry breaking we have to look
at terms that mix the singlet with the Higgses, as such terms 
are possibly also allowed by the symmetries.
These are invariants of the form
\be
\Theta ^\a H_u^\g H_d^\d \label{mix}
\ee
which will appear in the effective action for $\g + \d =0$ and 
$\a + \g q_X^{(H_u)} + \d q_X^{(H_d)} = 0$.
A term of the type
$\Theta ^\a$ which could break supersymmetry (too) strongly
is clearly not allowed by gauge invariance.  It is not easy to make general
statements about $F$-terms so let us be more specific and use the
D-brane inspired charge assignement of \cite {AKT} where
$X$ is identified with a Peccei-Quinn like symmetry and the Higgs charges are such that
$q_X^{(H_u)}-q_X^{(H_d)} = -4$.
The lowest dimension mixed invariant in the superpotential is then
$\frac{1}{M^3} \Theta ^4 H_u^\dagger H_d$.
The $F$-terms associated with the Higgs fields receive contributions from many
other terms besides (\ref{mix}) so it is a model dependent question whether
and how strongly they break supersymmetry. 
We can comment though on $F_{\Theta} \sim \frac{\Theta^3}{M^3} H_u^\dagger H_d$
which breaks supersymmetry by generating the 
rather soft term $|F_{\Theta}|^2\sim \l_c^6 |H_u^\dagger H_d|^2$
in the $V$-vacum. 
Clearly, other sources of supersymmetry breaking could be present,
in particular there could be $D$-term contributions along the lines suggested by \cite{BD}.
Since we would like to keep our analysis as general as possible we will not 
specify a supersymmetry breaking scenario we will instead proceed in
the spirit of the MSSM and add the most general supersymmetry breaking
terms to the Lagrangean. 

{\bf The Higgs mechanism}. The scalar potential should not only 
break supersymmetry but also EW symmetry. It includes the terms
\be
{\cal V} = D_X^2 + D_Y^2 + D_{SU(2)}^2 + \sum_\Phi |F_\Phi|^2 + m_{H_u}^2 H_u^\dagger H_u +
m_{H_d}^2 H_d^\dagger H_d.
\ee
This is a potential of the  form analyzed in detail in \cite{CIK} and it consists of the standard form
\bea
V &=& B H_u^\dagger H_d + c.c.
+ \m_1^2H_u^\dagger H_u + \m_2 ^2 H_d^\dagger H_d\nonumber\\
&+&\l_u(H_u^\dagger H_u)^2+\l_d(H_d^\dagger H_d)^2\nonumber\\
&-&\l_{ud}(H_u^\dagger H_u)(H_d^\dagger H_d) 
+\l_{ud}'\left|H_u^T\tau^2H_d\right|^2.\label{MSSMpotD}
\eea
plus the unconventional Higgs-axion mixing terms
\begin{eqnarray}
V^\prime &=&\l_0\, (H_u^{\dagger}H_d e^{-i\sum_I(q_I^{(H_u)}-q_I^{(H_d)})\frac{\theta_I}{M_I}})\nonumber\\
&+& \l_1 (H_u^{\dagger}H_de^{-i\sum_I(q_I^{(H_u)}-q_I^{(H_d)})\frac{\theta_I}{M_I}})^2  \nonumber\\
&+& \l_2 (H_u^{\dagger}H_u)(H_u^{\dagger}H_de^{-i\sum_I(q_I^{(H_u)}-q_I^{(H_d)})\frac{\theta_I}{M_I}})\nonumber\\
&+& \l_3 (H_d^{\dagger}H_d)(H_u^{\dagger}H_de^{-i\sum_I(q_I^{(H_u)}-q_I^{(H_d)})\frac{\theta_I}{M_I}}) + c.c.
\nonumber\\ \label{PQbreak}
\end{eqnarray}
where we have generalized to an arbitrary number of anomalous $U(1)_I$ gauge factors.
Notice that since the axions $\theta_I$ shift under $U(1)_I$ as
\be
\theta_I \longrightarrow \theta_I + \e_I M_I
\ee
the potential $V'$ is an allowed invariant, even though it does not seem to have 
an obvious $D$ or $F$-term origin. It could originate though from instanton effects
or from the decoupling of the heavy scalar singlet during supersymmetry breaking.
In the first case its contribution is expected to be tiny while in the second case
it could be of the order of the electroweak scale.
One can choose $q^{(\Theta_I)}_Y=0$ and then the necessary condition for the 
existence of a supersymmetric $V_I$-vacum  is that ${\rm det}({\bf A})\ne 0$ where $\bf A$ is
the square matrix with entries the anomalous charges of the $\Theta_I$ under $U(1)_J$ \cite{ILR}. 

In the $v$-vacum the original $V_I$-vacum will shift by a small amount
\be
<|\Theta_I|> = V_I + {\cal O}(v/M_I)
\ee
with $v^2=v_u^2+v_d^2$
and the rest of the Lagrangean can be consistently written in the new broken phase.
In \cite{CIK} it was shown that the potential ${\cal V}=V+V^\prime$ can break EW symmetry 
and give masses to the $Z, W^\pm$ gauge bosons while leaving the photon massless.
The masses of the $Z, W^\pm$ are corrected with respect to their SM vaues by effects
of order ${\cal O}(v/M)$, which can be used to give a lower bound on $M$ \cite{GIIQ,G1}.
The masses of the heavy gauge bosons are of order ${\cal O}(M)$.
In the scalar sector one obtains a phenomenology similar to that of the MSSM, with
the exception of the CP-odd state which mixes with the phases $\theta_I$ and results
in a physical CP-odd scalar, the axi-Higgs $\chi$, whose mass comes exclusively from
terms appearing in $V^\prime$ \cite{CIK}:
\be
m_\chi^2 \sim \sum_i c_i \l_i v^2 + {\rm small \; corrections}\label{axiHiggs}
\ee
with $c_i$ known coefficients (not too small) and $\l_i, i=0,1,2,3$ the couplings in $V^\prime$. 
The axi-Higgs decays through the interaction
\be 
g^{\chi F F} \chi\; F \wedge F
\ee 
which couples the axi-Higgs $\chi$ and two gauge bosons of the spontaneously broken phase 
and through the 1-loop fermion triangle which is of the same order of magnitude \cite{CIK, CIM1}. 
It is worth noting that while $g^{\chi F F}$ is suppressed,
$\chi$ couples rather strongly
\be
g^{\chi {\overline\psi}_u \psi_u}\sim N \cos\beta\hskip 1cm g^{\chi {\overline\psi}_d \psi_d}\sim N \sin\beta
\ee
to the up and down type of fermions respectively (dropping the flavor structure here) and
where $N=1+{\rm small\, corrections}$, and $\tan \beta=v_u/v_d$ the usual MSSM parameter.
An interesting feature we find here is that the mass and the coupling of $\chi$
to the gauge bosons and fermions are rather loosely connected, 
as opposed to conventional axion models where all three parameters
(mass, coupling to gauge bosons, coupling to fermions)
are governed by the same scale $f_a$.

{\bf Gauge Invariance $\&$ Unitarity.} 
In the analysis of the full Lagrangean one should in principle keep all fields,
including the superpartners. If however these theories are to be used as valid extensions of the SM,
there must exist a region in the parameter space
(at low enough energies) where the superpartners can be safely decoupled.
This does not hold though for the $Z'$ gauge boson and the axi-Higgs,
even at energy scales $E\sim v$.
This peculiarity originates from the scale blind nature of the anomaly which requires 
a specific minimal structure in the low energy effective Lagrangean.
This structure we called minimal low scale orientifold models (MLSOM) in \cite{CIK}
and it has the form (shown here for a single extra $U(1)$; 
the generalization to an arbitrary number of anomalous $U(1)$'s is straightforward, see \cite{CIK})
\be
{\cal L} = {\cal L}^{SM}(H_u, H_d) + (\partial_\m \theta + M A^X_\m)^2 + 
g^{\theta F{\tilde F}}\frac{1}{M} \;\theta F \wedge F +
{\cal L}^{anom}
\ee
where the first part is just the Standard Model Lagrangean but with two Higgs doublets
instead of one and the rest of the terms represent the Green-Schwarz structure necessary
for the consistency of the quantum effective action. 
The last piece in particular, the one loop triangle anomaly contribution,
needs special care since there are possible ambiguities hiding in the definition
of triangles of non-vanishing traces. A detailed study of these ambiguities can be
found in \cite{ABDK, CIM1}. Here we just mention one interesting feature \cite{CIM1}, namely that 
unitarity at one-loop requires that the Higgs fields be charged under all anomalous $U(1)$'s
($q_X^{(H_u)}, q_X^{(H_d)}\ne 0$), 
a fact that is not dictated by gauge invariance.

{\bf Phenomenology}.
Given that one expects to see this structure in any string compactification 
that includes the Standard Model at low energies (almost any such construction up to date
in either the heterotic or the open string context has anomalous $U(1)$'s and thus the whole
package that comes with them)
the obvious question is if one can distinguish such a model from other extensions of the SM.
To begin, anomalous $U(1)$ models have naturally 
a number of $Z'$ gauge bosons that contribute significantly
to Drell-Yan processes (even though extra non-anomalous $U(1)$'s have similar effects).   
The process though that definitely is a signature of anomalous $U(1)$ models is the decay
of a particle through the fermion triangle. In the SM these contributions 
are different or even vanish identically
in the massless fermion limit due to the anomaly free nature of hypercharge.
Here, since there are non-vanishing traces there will be non-zero contributions 
even in the massless fermion limit. Two characteristic examples are the decay of the axi-Higgs
into two photons $\chi \longrightarrow \g\g$ 
and the decay $Z^*\longrightarrow \g\g$, an off-shell $Z$ going into two photons. 
The first, in the case where $\chi$ is EW scale
heavy (the $\l_i$ in (\ref{axiHiggs}) are not too small) is a contribution to the decay of a
Higgs like particle into two photons and even though such a state does not exist in the SM,
it could be observed at the LHC together with similar decays of the usual Higgs particles.
We remind that the property that distinguishes $\chi$ from a genuine Higgs, 
the CP-odd MSSM Higgs $A^0$ in our case, is its direct coupling to the gauge bosons.  
The second is a process that in the SM receives contributions only from massive fermions
but it has a non-trivial piece also from the zero fermion mass limit in the anomalous $U(1)$ extensions
due to the non-vanishing traces. Thus, this is a process which is clearly comparable to the 
corresponding one in the SM and in addition it is a characteristic signature of these models.
Quite generally, an experimental indication corresponding to the "decay" of a $Z^*$ into two photons
larger than the SM prediction, is an indirect evidence for an anomalous structure.
A further complication stems form the fact that the $Z^*$ makes sense only as an intermediate
state (by the Landau-Yang theorem) and thus one should look for the new signal 
for example in the process $q + {\overline q}\longrightarrow \g\g$
with an intermediate gauge boson or a scalar exchanged either directly 
($\chi^*$) or through a fermion triangle ($\g^*$, $Z^*$, $Z^{\prime *}$, $\chi^*$). 
There are of course other contributions but
it is imoprtant not to forget that the above
channels are linked via the anomalous Ward identities \cite{CIM1}.

Finally, in the case where $\l_i<<1$ the axi-Higgs is closer in nature to a conventional PQ axion
and therefore its phenomenology should be analyzed in that context
(here astrophysical and cosmological considerations become important).
Such an analysis will be performed elsewhere with a focus on the recent PVLAS data.

\centerline{\bf Acknowledgements}
We would like to thank E. Kiritsis for collaboration in parts of this project and T. Tomaras for
discussions.
The work of C.C. is partially supported by INFN of Italy
(grant BA21).  

\newpage



\end{document}